\documentclass[
    ,final            
  ]
  {aipproc}

\layoutstyle{6x9}

\begin{document}
\newcommand{\pT}{$p_{\mathrm{T}}$}
\newcommand{\ifb}{$\mathrm{fb^{-1}}$}
\newcommand{\sqrts}{$\sqrt{s}$} 
\newcommand{\lumi}{$\mathrm{cm^{-2}s^{-1}}$} 
\newcommand{\Bspsiphi}{$\mathrm{B^0_s \to J/\psi\phi}$}
\newcommand{\Bppsik}{$\mathrm{B^+ \to J/\psi K^+}$}
\newcommand{\Bstomumu}{$\mathrm{B^0_s \to \mu^+\mu^-}$}
\newcommand{\Bdtomumu}{$\mathrm{B^0 \to \mu^+\mu^-}$}
\newcommand{\Btomumu}{$\mathrm{B^0_{(s)} \to \mu^+\mu^-}$}
\newcommand{\Bd}{$\mathrm{B^0}$}
\newcommand{\Bp}{$\mathrm{B^+}$}
\newcommand{\Bs}{$\mathrm{B^0_s}$}

\title{Rare \Btomumu decays}


\author{P. Eerola \\ for the ATLAS, CDF, $\underline{CMS}$, D0 and LHCb collaborations}{
  address={University of Helsinki and Helsinki Institute of Physics}
}

\begin{abstract}
 This review summarizes the current experimental results on rare \Btomumu\ decays of the Tevatron experiments CDF and D0, and the
 LHC experiments ATLAS, CMS and LHCb. The experimental branching fraction upper limits for the \Bstomumu\ are already quite close to the branching fraction predicted by
 the Standard Model, and the first observation of the \Bstomumu decay is expected soon. The rare decays \Btomumu\ are highly suppressed in the Standard Model, and
 therefore accurate measurements of these branching fractions provide complementary constraints to the free parameters of various extensions of
 the Standard Model.
 
 \end{abstract}

\maketitle


\section{Introduction}

The decay channels \Btomumu\ are flavour-changing neutral current processes, which are forbidden 
in the Standard Model (SM) at the tree level, and occur only via higher order diagrams. In addition, the decays
are helicity suppressed by $m^2_{\mu}/m^2_{\mathrm{B}}$. The branching fraction predicted by the SM for the decay channel \Bstomumu\ is 
$(3.2 \pm 0.2) \cdot 10^{-9}$, and the predicted SM branching fraction for the decay \Bdtomumu\ is  $(1.0 \pm 0.1) \cdot 10^{-10}$ \cite{Bsbr}. 
Possible new particles contributing in higher order diagrams 
typically increase the branching fractions, and therefore these decay channels are 
sensitive to new physics beyond the SM. The branching fractions can be, however, also suppressed with respect to the SM predicted values in some regions of the parameter space \cite{suppr}. Precision measurements of the branching fractions and their ratio 
will thus serve as a test of the SM and as a constraint of the allowed parameter space of models beyond the SM. For a recent theoretical overview on the constraints on new physics in rare B decays, see $e.g.$ Ref.~\cite{straub}. 

It has recently been pointed out \cite{debruyn}, that the SM prediction for the \Bstomumu\ branching fraction becomes $(3.5 \pm 0.2) \cdot 10^{-9}$,
when taking into account that the \Bstomumu\ decays are 
integrated over all decay times and averaged over untagged $\mathrm{B^0_s}$ decays. 
The SM fit of the CKM matrix and CP violation parameters \cite{CKMFitter} 
gives BR(\Bstomumu)=$(3.64^{+0.21}_{-0.32}) \cdot 10^{-9}$.

The branching fraction BR(\Bstomumu) (BR(\Bdtomumu)) can be measured relative to a well-known decay, for example \Bppsik, as:
 \begin{equation}
BR(\mathrm{B^0_s \to \mu^+\mu^-}) =  \frac{N^{signal}_{obs}}{N^{B^+}_{obs}}   \frac{f_u}{f_s} \frac{\varepsilon^{B^+}_{tot}}{\varepsilon^{signal}_{tot}} 
BR(\mathrm{B^+  \to J/\psi K^+ \to \mu^+\mu^- K^+}),
\end{equation}
where $ N^{signal(B^+)}_{obs}$ is the number of observed signal ($\mathrm{B^+}$) events, $\varepsilon^{signal(B^+)}_{tot}$ the total efficiency to trigger and reconstruct signal ($\mathrm{B^+}$) events, and $f_u / f_s$ is the fraction of the production fraction of $\mathrm{B^+_{(u)}}$ and $\mathrm{B^0_s}$ mesons. 

Experimentally the channels \Btomumu\ are relatively simple to trigger and reconstruct. The combinatorial backgrounds  $\mathrm{b\bar{b} \to \mu^+\mu^-X}$ and 
$\mathrm{b \to c\mu^-\nu \to  s\mu^+\mu^-\nu\nu  }$ constitute  
typically the major background for \Bstomumu. These background sources can be measured from the data by fitting the signal mass region sidebands. Peaking backgrounds $\mathrm{B \to h^-\mu^+\nu}$ and $\mathrm{B \to h^+h^-}$, where $\mathrm{h = \pi, K, p}$ are misidentified as muons, are more important for \Bdtomumu\ because of the lower mass 
and smaller branching fraction, while for \Bstomumu\ they play a minor role. CDF, CMS and LHCb experiments have sufficient mass resolution to clearly separate
the \Bd\ and \Bs\ mass peaks. Muon misidentification rates can be derived from the data for example by using $\mathrm{D^*}$-tagged $\mathrm{D^0 \to K^-\pi^+}$ decays. The backgrounds from other exclusive decays such as $\mathrm{B^0_s \to \mu^+\mu^-\gamma}$, $\mathrm{B^+ \to \mu^+\mu^-\pi^+}$ and $\mathrm{B_c \to J/\psi\mu^+\nu}$ are so far negligible.

Typical discriminating variables used in analyses are transverse momenta $p_{\mathrm T}$ of the B and of the muons, quality of the secondary (dimuon) and primary vertex fits,
requirements on the secondary dimuon vertex (flight length and significance, impact parameter of the B and/or of the muons and their significance), pointing angle of the B,
isolation of the dimuon, and association of the B to the correct primary vertex.  The experiments perform their analyses blinded, $i.e.$ the selection cuts are tuned without looking at the signal mass region.

\section{Tevatron results from the D0 and CDF experiments}

The Tevatron operation was terminated in September 2011, after having delivered over 10~\ifb\ for the D0 and CDF experiments. 
The peak instantaneous luminosity achieved was L = $ \mathrm{4.3 \cdot 10^{32} cm^{-2} s^{-1}}$.

The results by the D0 experiment were obtained with an integrated luminosity of 6.1~\ifb \cite{d0results}.
The data were collected with a set of dimuon and single muon triggers, and energy depositions in the calorimeter.
A Bayesian Neural Network, trained with simulated signal events 
and data from mass sidebands, was used to obtain the branching fraction limit
BR(\Bstomumu) $<51 \cdot 10^{-9}$ at 95\% CL. The median expected limit was BR(\Bstomumu) $<40 \cdot 10^{-9}$ at 95\% CL.
Peaking backgrounds were estimated from simulation and from semileptonic B decays. The D0 dimuon mass resolution was 120 MeV,
and therefore \Bs\ and \Bd\ decays could not be separated. The result was given for \Bs\ only, assuming no background from \Bdtomumu.


The CDF preliminary results were obtained with an integrated luminosity of 10~\ifb \cite{cdfresults}.
The data were collected with dimuon triggers and requiring that at least one of the muons is in the central detector region.
Hadrons misidentified as muons were suppressed by using a likelihood method
combined with a $dE/dx$ based selection.

A Neural Network (NN) was used to separate signal and background. The NN was the same as in the previous analysis with 
7~\ifb~\cite{cdfprev}, without any retraining.  CDF found a mild excess of events in the highest NN output bins of the \Bstomumu\ search ($\nu_{NN} > 0.97$), 
shown in Fig.~\ref{fig2}. Using a log-likelihood fit CDF reported BR(\Bstomumu) $ = (13^{+9}_{-7}) \cdot 10^{-9} $,
where the central value corresponds to the minimum of the $\Delta \chi^2$ distribution, and the uncertainty was taken as one unit of change in the $\Delta \chi^2$.
The result corresponds to the branching fraction limit BR(\Bstomumu) $<31 \cdot 10^{-9}$ at 95\% CL (median expected limit BR(\Bstomumu) $<13 \cdot 10^{-9}$ at 95\% CL). In the \Bdtomumu\ search the data were consistent with the background-only expectations. The obtained limit was
BR(\Bdtomumu) $<4.6 \cdot 10^{-9}$ at 95\% CL (median expected limit BR(\Bdtomumu) $<4.2 \cdot 10^{-9}$ at 95\% CL).

\begin{figure}[htb]
  \includegraphics[height=.3\textheight]{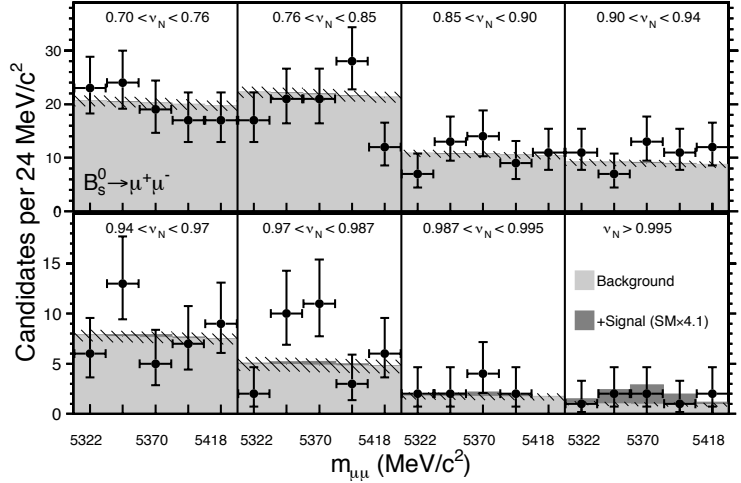}
  \caption{CDF dimuon mass distributions for different NN output ($\nu_{NN}$) bins. The expected SM $\mathrm{B^0_s \to \mu^+\mu^-}$ signal has been scaled with a factor of 4.1.}
  \label{fig2}
\end{figure}

\section{LHC results from the ATLAS, CMS and LHCb experiments}
The experimental results from LHC reported here stem from the data collected in 2011. ATLAS and CMS experiments collected an integrated luminosity of about 5~\ifb. LHCb collected an integrated luminosity of about 1~\ifb, because the instantaneous luminosity was leveled to a maximum of about $\mathrm{4 \cdot 10^{32} cm^{-2}s^{-1}}$.

The ATLAS experiment~\cite{atlas} results on the \Bstomumu branching fraction were obtained with an integrated luminosity 
of 2.4 \ifb~\cite{atlasbstomumu}. Mainly dimuon triggers were used to collect the data. The analysis was performed 
separately for three different pseudorapidity regions $|\eta| <1.0$, $1.0 < |\eta| <1.5$, and $1.5 < |\eta| <2.5$,
because of the different mass resolutions in these regions. Boosted decision trees (BDT) were used to separate signal
and background. ATLAS has no sensitivity to \Bdtomumu, because the mass resolution is not sufficient to separate 
\Bs\ and \Bd\ decays. The background from \Bdtomumu\ decay to the \Bstomumu\ channel was assumed to be zero. ATLAS obtained a branching fraction limit of
BR(\Bstomumu) $<22 \cdot 10^{-9}$ at 95\% CL. 
The median expected limit was BR(\Bstomumu) $<25 \cdot 10^{-9}$ at 95\% CL.

The results by the CMS experiment~\cite{cms} were obtained with an integrated luminosity of 5~\ifb~\cite{bstomumu}.
The dataset was collected with a displaced dimuon trigger, requiring $4.8 < m_{\mu\mu} < 6.0$~GeV and that the muons originate from a common good-quality
three-dimensionally measured vertex. The cut-and-count analysis was tuned separately for the barrel (both muons with $|\eta|<1.4$) and endcap 
(at least one of the muons with $|\eta|>1.4$) regions.  Pileup independence of the selection efficiences was carefully studied with the \Bppsik\ and \Bspsiphi\ data samples. The selection efficiencies were found to be constant for events with up to 30 primary vertices. The branching fraction limits obtained were:
BR(\Bstomumu) $<7.7 \cdot 10^{-9}$ at 95\% CL (median expected limit BR(\Bstomumu) $<8.4\cdot 10^{-9}$ at 95\% CL), and
BR(\Bstomumu) $<1.8 \cdot 10^{-9}$ at 95\% CL (median expected limit BR(\Bstomumu) $<1.6\cdot 10^{-9}$ at 95\% CL). 
The final dimuon mass distributions are shown in Fig.~\ref{fig3}.

\begin{figure}[htb]
 \includegraphics[height=.25\textheight]{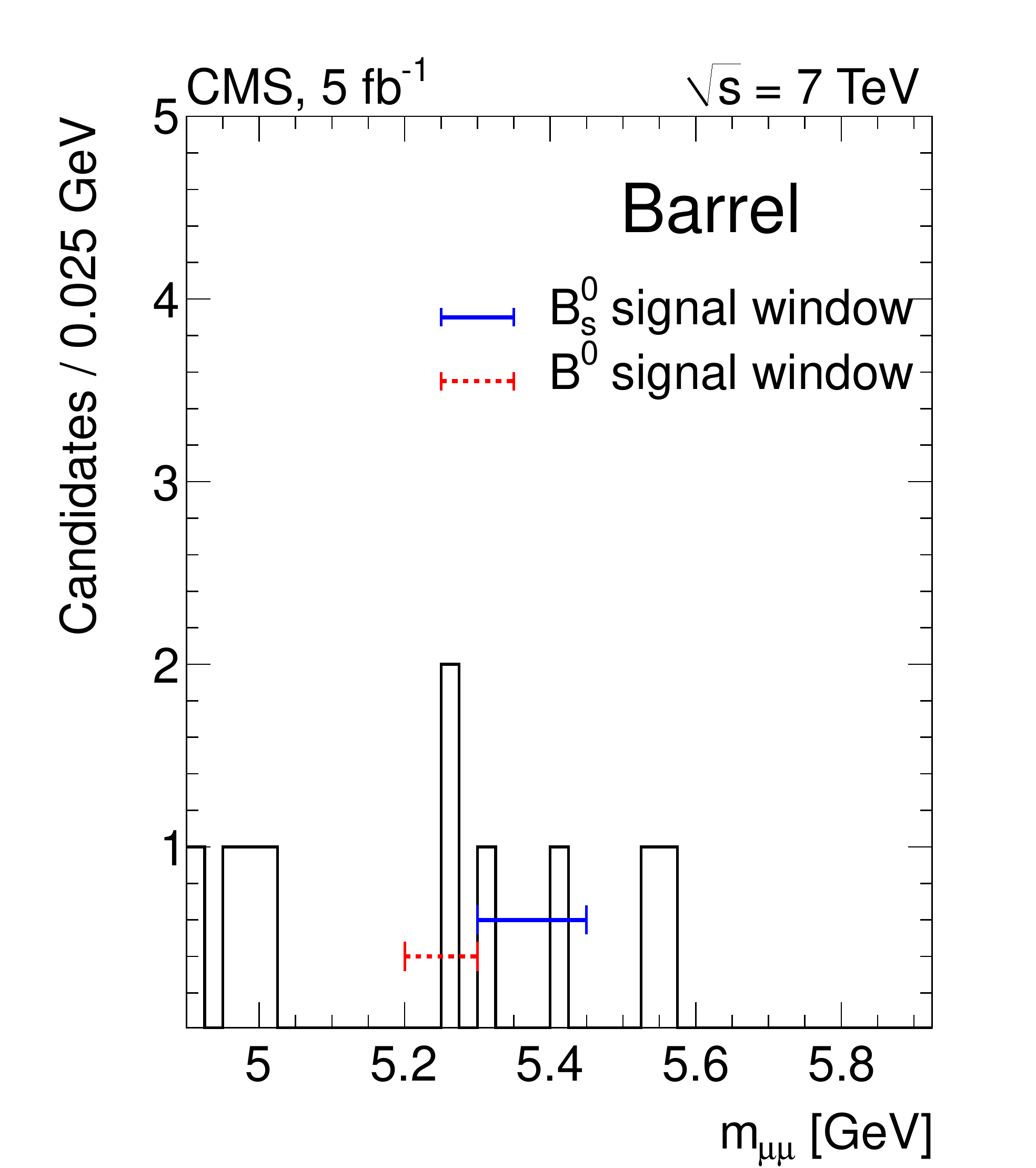}
 \includegraphics[height=.25\textheight]{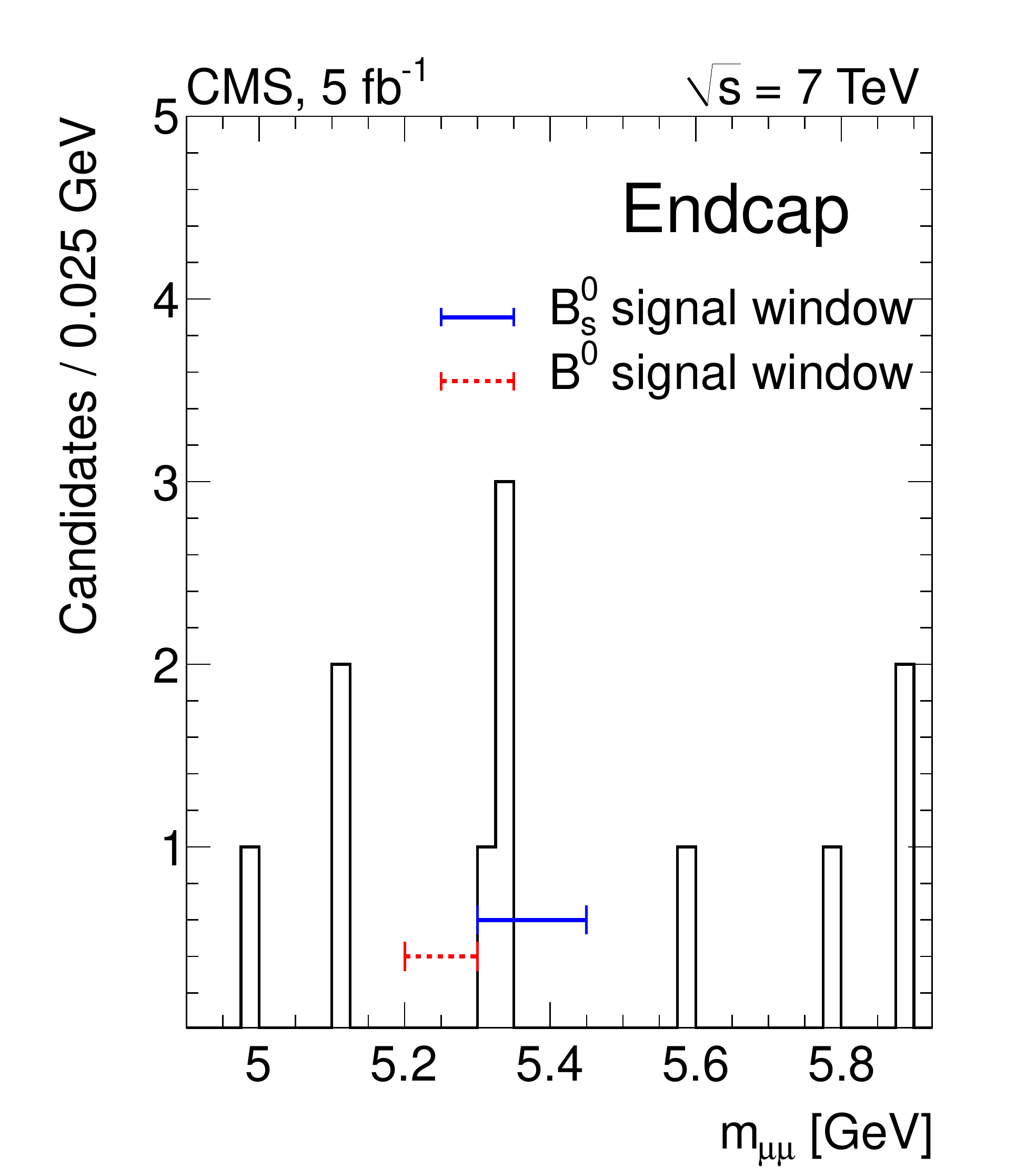}
  \caption{CMS dimuon mass distributions. Left: barrel region, right: endcap region.}
\label{fig3}
\end{figure}

The LHCb experiment~\cite{lhcb} used an integrated luminosity of 1~\ifb\ to obtain their \Btomumu\ results~\cite{lhcbbstomumu}.
LHCb used three different normalization channels: \Bppsik, \Bspsiphi, and $\mathrm{B^0 \to K^+\pi^-}$. Charged hadron identification by the RICH detectors was used to separate kaons, pions and protons. The decays $\mathrm{B^0 \to h^+h^-}$, $\mathrm{D^0 \to K^-\pi^+}$, and dimuon resonances were used to control the particle identification, mass scale, and multivariate methods.

The analysis was performed in two stages with two different BDTs. The obtained branching fraction limit for the  \Bstomumu\ was 
BR(\Bstomumu) $<4.5 \cdot 10^{-9}$ at 95\% CL (median expected upper limit for background and SM signal prediction $7.2 \cdot 10^{-9}$ at 95\% CL). Using an unbinned maximum likelihood fit to the mass projections of the BDT output bins, the branching fraction was found 
to be BR(\Bstomumu)~$ = (0.8^{+1.8}_{-1.3}) \cdot 10^{-9}$.
The obtained limit for the \Bdtomumu\ was BR(\Bdtomumu) $<1.03 \cdot 10^{-9}$ at 95\% CL (median expected upper limit $1.13 \cdot 10^{-9}$ at 95\% CL). 
The dimuon mass distributions summed over BDT bins with output greater than 0.5 are shown in Fig.~\ref{fig4}.

\begin{figure}[htb]
 \includegraphics[height=.25\textheight]{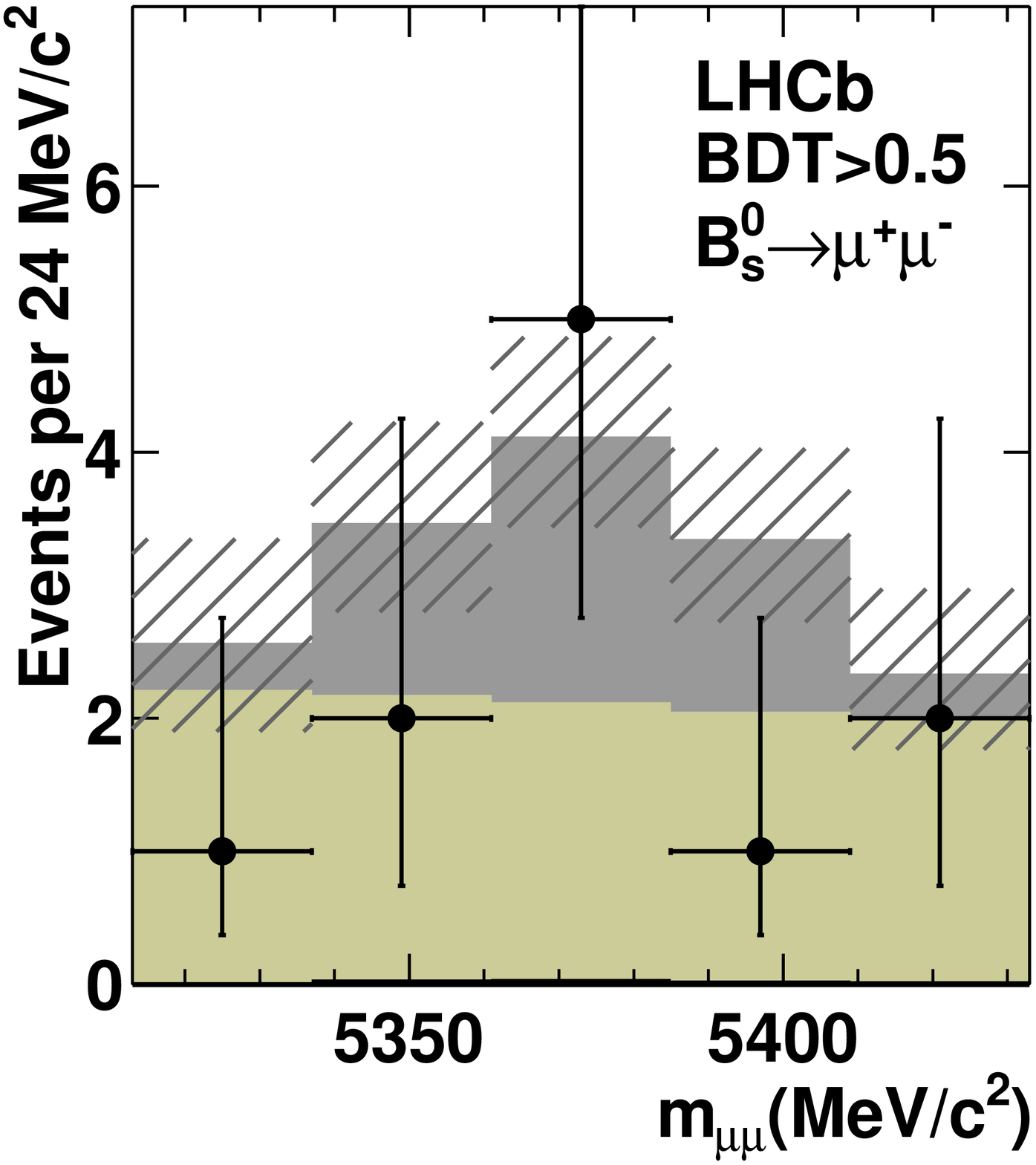}
 \includegraphics[height=.25\textheight]{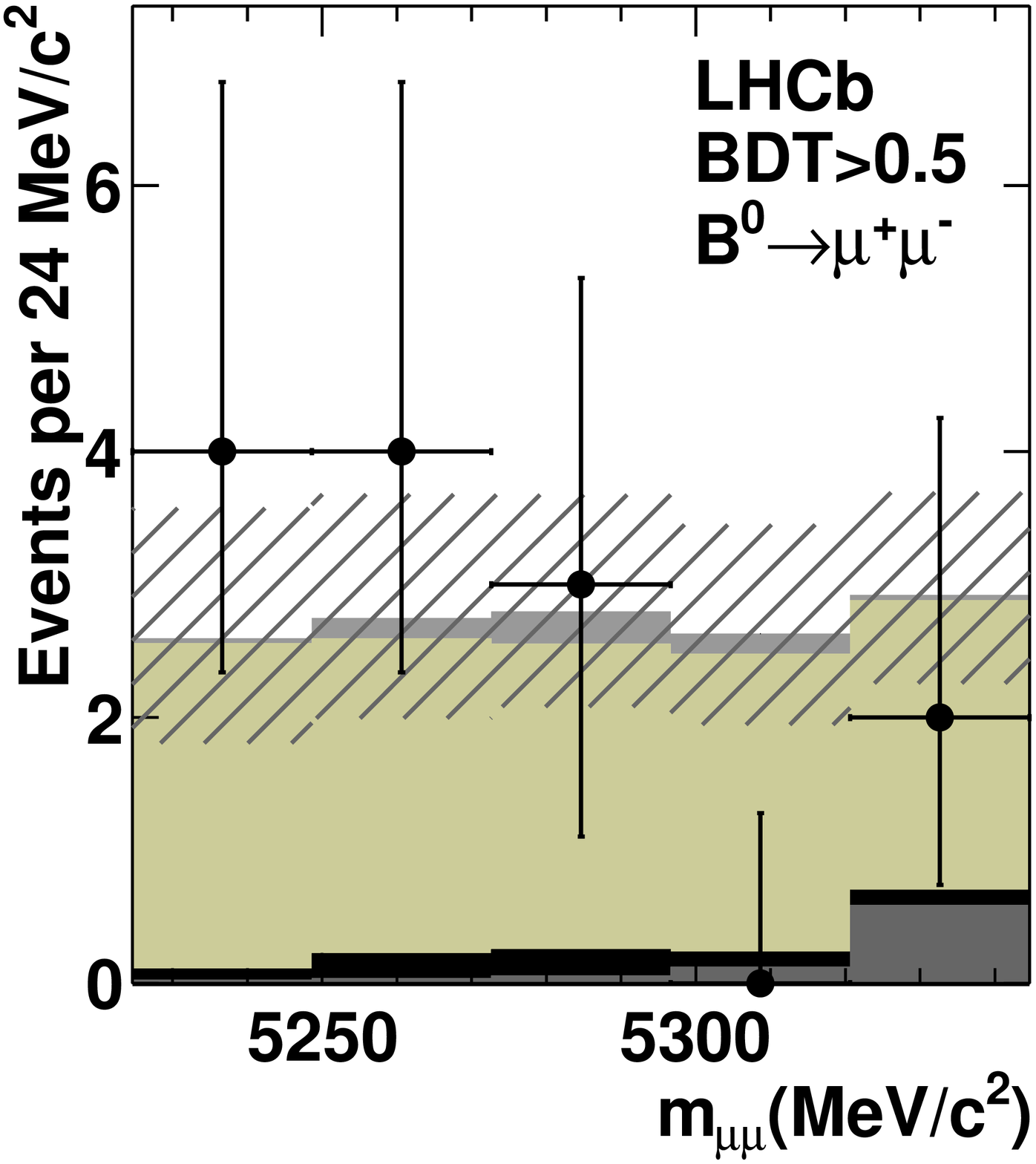}

  \caption{LHCb dimuon mass distributions, summed over BDT bins with output greater than 0.5. Left: $\mathrm{B^0_s \to \mu^+\mu^-}$ signal mass region, right: $\mathrm{B^0 \to \mu^+\mu^-}$ signal mass  region.}
\label{fig4}
\end{figure}

The combined result of ATLAS, CMS and LHCb for the \Bstomumu decay channel is currently BR(\Bstomumu) $<4.2 \cdot 10^{-9}$ at 95\% CL 
\cite{bstomumucomb}. This result constrains large values of $\tan\beta$, and is
complementary to the direct searches. The combined result of CMS and LHCb for the \Bdtomumu decay channel is 
BR(\Bdtomumu) $<8.1 \cdot 10^{-10}$ at 95\% CL \cite{bstomumucomb}. 

\section{Prospects and outlook}

The CMS and LHCb experiments have good chances for the first observation of \Bstomumu\ in 2012, if the branching fraction is in the range predicted by the SM. 
In 2012 it is expected that ATLAS and CMS experiments will collect an integrated luminosity of at least 20~\ifb\ each, on top of the 5~\ifb\ collected in 2011. LHCb will collect an integrated luminosity of 1.5 -- 2~\ifb,  to be added to the 1~\ifb\ collected in 2011.
It is unlikely that the decay \Bdtomumu\ would be observed 
at the LHC any time soon, without very significant enhancements to the SM rate. 
Nevertheless, searches for \Bdtomumu\ will be pursued as a long-term goal.




\bibliographystyle{aipproc}   




\end{document}

\endinput


\begin{table}
\begin{tabular}{lrrrr}
\hline
  & \tablehead{1}{r}{b}{Single\\outlet}
  & \tablehead{1}{r}{b}{Small\tablenote{2-9 retail outlets}\\multiple}
  & \tablehead{1}{r}{b}{Large\\multiple}
  & \tablehead{1}{r}{b}{Total}   \\
\hline
1982 & 98 & 129 & 620    & 847\\
1987 & 138 & 176 & 1000  & 1314\\
1991 & 173 & 248 & 1230  & 1651\\
1998\tablenote{predicted} & 200 & 300 & 1500  & 2000\\
\hline
\end{tabular}
\caption{Average turnover per shop: by type
  of retail organisation}
\label{tab:a}
\end{table}